\documentclass[12pt]{article}
\usepackage{amsfonts} 
\usepackage[utf8]{inputenc}
\usepackage{amssymb, amsmath}
\usepackage{color}
\usepackage{cite}

\title{Angular momentum in the internal SO(3) symmetry  space relating the size of universe and the Higgs mass}

\author{
Metin Arık \footnote{metin.arik@boun.edu.tr} ,  Tarik Tok \footnote{tarik.tok@boun.edu.tr}\\
  Department of Physics,
 Bogazici University,
   Bebek, Istanbul, Turkey }

\begin{document}
\maketitle
\begin{abstract}
We show that with an internal SO(3) symmetric triplet of Higgs fields,  the conserved quantity associated with this internal SO(3) symmetry leads to spontaneous symmetry breaking  giving the Higgs field a mass.
\end{abstract}

The Standard Model is perhaps the most extraordinary achievement of particle physics. It not only described the electromagnetic, weak, and strong interactions but also categorized elementary particles. Until the discovery of the Higgs boson by the ATLAS\cite{Atlas} and CMS\cite{CMS} experiments at CERN's large hadron collider in 2012, the theoretical framework of particle physics had been contradicted with the experimental result of non-zero particle masses, so this theory needed an extension. François Englert\cite{englert} and Peter Higgs\cite{higgs} contributed to the concept of the origin of the mass of subatomic particles. This discovery is named as the Higgs Mechanism, which solved the problem, giving mass to W\&Z bosons and fermions by using their interaction with an invisible field, called the “Higgs field”, that permeates the universe. This quantum field generates spontaneous symmetry breaking within self-interactions. This symmetry breaking triggers the Higgs mechanism.

Abdus Salam and Steven Weinberg established the Higgs mechanism as an essential part of particle physics. When electroweak symmetry is unbroken, all elementary particles are massless in the Standard Model. The mass is a fundamental result of the vacuum expectation value of the Higgs field. The symmetry is spontaneously broken by tachyon condensation, and the $W^{+}$, $W^{-}$ and Z bosons acquire masses.\cite{higgs, kondo}.

 In the Standard Model, the Higgs field plays a crucial role in cosmology\cite{royalsociety, bosoncosmology}, leading to the Universe's flatness, homogeneity, and isotropy, creating the quantum fluctuations, triggering the radiation-dominated era of the Big Bang\cite{bigbang}, and making contribution to the production of dark matter\cite{darkmatter1} and baryogenesis\cite{baryon}. Besides, spontaneous symmetry breaks with tachyonic condensation \cite{d'Onofrio,tacyon2}, the Higgs mechanism also appears in topologically massive gauge theories in an arbitrary space-time dimension\cite{ichiroodatopology1}. It has also been remarked that the Higgs field can create its own geometry\cite{geo,geo2}.
And also, the Standard Model Higgs field non-minimally coupled to gravity\cite{hooft, mukhanov, odagraviton} could be responsible for the Palatini-Higgs inflation scenario \cite{palatinihiggs, palatini2}
.In addition to these, in the first-quantized string theory,  Goldstone bosons can be found after SU(2) symmetry breaking\cite{string}. 
To sum up, the Higgs mechanism attracted a lot of attention in various disciplines of physics because of its elegance.

Although the Higgs field is an object in quantum field theory, initially,  we handle the vacuum expectation value of the Higgs field as a classical field which depends on time due to the isotropy and homoneity of the Universe.  After first quantization, spontaneous symmetry breaking will be obtained. 

In the Standard Model, the Higgs field,  by adding a tachyonic negative mass squared term to the Lagrangian density,  induces spontaneous symmetry breaking.  On the other hand,  in this paper,  we will show that symmetry breaking can occur without a negative mass squared term\cite{paper}. Note that  SU(2) and SO(3) are locally isomorphic \cite{hall}. We consider the simplest non-Abelian Higgs model with the SO(3) symmetry. Thus,  in our model, the new quantum number $\mathsf{j}$  associated with the internal SO(3) invariance of the field space is conserved and may lead to spontaneous symmetry breaking which results in a Higgs mass inversely proportional to the size of the universe. Such a  mass can be useful in cosmological models and in explaining the neutrino mass.

The Minkowski metric is defined as

$$\eta_{\mu\nu}=(+1,-1,-1,-1)$$
and we use units where $\hbar=c=1$.

The Lagrangian density of the Higgs field 

\begin{equation}
\mathcal{L}= \partial_\mu \Theta^\dagger \partial^\mu \Theta -V(\Theta)
\end{equation}
with 
$
\Theta=\begin{pmatrix} 
\phi_1 \\
\phi_2 \\
\phi_3
\end{pmatrix}
$
where $\phi_a$ corresponds to scalar fields and  $a=1,2,3$. Furthermore, the potential term is usually  defined as

\begin{equation}
V(\Theta)=-\frac{1}{2}\bar{m}^2\Theta^\dagger\Theta+\frac{\lambda}{4}(\Theta^\dagger\Theta)^2
\end{equation}
where dimensionless constant $\lambda \textgreater  0$ and the scalar fields have the dimension of mass.

In terms of the fields $\phi_a$, the Lagrangian density can be written as
\begin{equation}\label{eql}
\mathcal{L}= \frac{1}{2}\partial_\mu\phi_a(t,\vec{r})\partial^\mu\phi_a(t,\vec{r})-\frac{\lambda}{4} (\phi_a(t,\vec{r})\phi_a(t,\vec{r}))^2
\end{equation}
where, we put $\bar{m}=0$.  

The system can be quantized by imposing the commutation relation

\begin{equation} \label{eq1}
\left[ \dot{\phi_a(t,\vec{r})},\phi_b(t,\vec{r'})\right]= -i \delta^3(\vec{r}-\vec{r'})\delta_{ab}.
\end{equation}
By the way, in homogeneous and isotropic cosmology, $\phi$ may depend on only time 

\begin{equation}
\phi(t,\vec{r})=\phi(t).
\end{equation}
We assume that the Universe is a torus that is flat, but finite, then the spatial volume of the universe is $a^3$. After integrating commutation relation overall space, the spatial volume of the universe appears

\begin{equation} \label{eq1}
\int\left[ \dot{\phi(t,\vec{r})},\phi(t,\vec{r'})\right] d^3r = -i \int \delta^3(\vec{r}-\vec{r'}) d^3r 
\end{equation}

\begin{equation} \label{eq1}
\left[ \dot{\phi(t)},\phi(t)\right] a^3 = -i 
\end{equation}
where a is the size of the universe. Actually, it is
changing with time, but we assume that a is constant in this paper.

The analogue of angular momentum density in the field space can be written as
\begin{equation} \label{eqn252}
\mathcal{J}_a=\epsilon_{abc} \phi_b \dot{\phi}_c
\end{equation}  
and, its dimension is $mass^3$.

We can plug $a^3$ into the commutation(\ref{eq1}) and distribute evenly

\begin{equation} 
\left[ \dot{\phi(t)}a^{\frac{3}{2}},\phi(t)a^{\frac{3}{2}}\right]  = -i 
\end{equation}
and one may define $\phi(t)a^{\frac{3}{2}}$ as a new field

\begin{equation}
\Phi(t)=a^{\frac{3}{2}} \phi(t)
\end{equation}
whose dimension is $mass^{-\frac{1}{2}}$.  We note that, in terms of new fields, the quantization of the system can be imposed by the commutation relation

\begin{equation} \label{eq2511}
\left[ \dot{\Phi}_a(t),\Phi_b(t)\right]= -i \delta_{ab}.
\end{equation}
By definition, the integral of the Lagrangian density over whole space gives the Lagrangian of the system

\begin{equation}
L=\int \mathcal{L} d^3 x.
\end{equation}
The Lagrangian density is independent of position variables because of isotropy$\&$homogeneity of the Universe, and the Lagrangian can be expressed as

\begin{equation}
L=  \mathcal{L} a^3.
\end{equation}
Therefore, we can investigate the system in the perspective of a SO(3) invariant internal space,

\begin{equation}
L= \frac{1}{2}\dot{\Phi}_a(t)\dot{\Phi}_a(t)-\frac{\Lambda}{4} (\Phi_a(t)\Phi_a(t))^2
\end{equation}
where $\Lambda$ is defined as
\begin{equation}
\Lambda= a^{-3} \lambda
\end{equation}
and, its dimension is $mass^3$.
Moreover, in the picture of the SO(3) symmetry, the analogue angular momentum in internal field space may be written as

\begin{equation} \label{eqn252}
J_a=\epsilon_{abc} \Phi_b \dot{\Phi}_c
\end{equation}
and, from (\ref{eq2511}), it satisfies

\begin{equation} \label{eq251}
\left[ J_a,J_b \right]= -i \epsilon_{abc} J_c
\end{equation}
which is the commutation relation of analogue angular momentum components.

After some algebraic calculations,  the Hamiltonian  becomes

\begin{equation}\label{4}
H=\frac{1}{2}\dot{\Phi}_a(t)\dot{\Phi}_a(t)+\frac{\Lambda}{4} (\Phi_a(t)\Phi_a(t))^2.
 \end{equation}
One can write this Hamiltonian in terms of  $\Phi=\sqrt{\Phi_1^2+\Phi_2^2+\Phi_3^2}$  with  $\Phi_a$ in the space that has the SO(3) symmetry

\begin{equation}\label{4}
H=\frac{1}{2}\dot{\Phi}^2(t)+\frac{J^2}{2 \Phi^2(t)}+
\frac{\Lambda}{4} \Phi^4(t)
 \end{equation}
where J is the analogue angular momentum operator of the internal field space in the aspect of
the conserved quantity associated with the SO(3) symmetry and is dimensionless. Here, the effective potential for our dynamical system is given by

\begin{equation}
V_{eff}=\frac{J^2}{2 \Phi^2(t)} +\frac{\Lambda}{4}\Phi^4.
\end{equation}
where the eigenvalue of $J^2$ is $\mathsf{j}(\mathsf{j}+1)$. Then,  its minimum  is given by 

\begin{equation}
\Phi_0=\bigg(\frac {\mathsf{j}(\mathsf{j}+1)}{\Lambda}\bigg)^\frac{1}{6}.
\end{equation}
Furthermore,  the field, $\Phi$, may be expanded pertubatively  as

\begin{equation}\label{eq15}
\Phi=\Phi_{0} +\epsilon \big(A  e^{imt}+ A^*  e^{-imt}\big)+\mathcal{O}(\epsilon^2).
\end{equation}
Here, $A$ and $A^*$ are annihilation and creation operators for the time dependent background expectation value.  They satify the cannonacal commutation relation.  After performing a perturbative calculation neglecting higher powers of $\epsilon$, we find that for consistency the mass $m$ and frequency  in   (\ref{eq15}) should be given by 

\begin{equation}\label{eq16}
m=\sqrt{12} \lambda^{\frac{1}{3}} \frac{({\mathsf{j}}({\mathsf{j}}+1))^\frac{1}{3}}{a}.  
\end{equation}
Note that $\Lambda=a^{-3} \lambda$ and $\mathsf{j}$ has to be big enough to find non-trivial Higgs mass. In other words, there must be an internal analogue angular momentum quantum number of unprecedented magnitude to obtain the measured Higgs mass.

In the Standard Model, without spontaneous symmetry breaking the scalar particle mass is given by a ghost term which introduces a free parameter with the dimension of mass. In our case, this free parameter is replaced by the dimensionless quantum number of internal symmetry and the inverse size of the universe.  We have studied spontaneous symmetry breaking in the
perspective of the internal SO(3) symmetry. Thus, a massive field with mass m has been obtained by starting from the Lagrangian density(\ref{eql}) which is the SO(3) restriction of the Standard Model's scalar sector.  In fact,  with all other fields set to zero,  the Standard Model scalar sector has the SO(4) symmetry.  Thus our future work will be related to an internal SO(4) symmetric scalar field space which has also been considered in other contexts\cite{conclusion1, conclusion2, conclusion3}.  We also plan to consider the important case when the size of three-dimensional space is time-dependent.

\bibliography{references}

\end{document}